\documentclass[prb,a4paper,twocolumn,showpacs]{revtex4-1}
	
    \usepackage[utf8]{inputenc}
    \usepackage[T1]{fontenc}

    \usepackage{amssymb}
    \usepackage{amsmath}
    \usepackage{amsbsy}
    \usepackage{bm}
    \usepackage{graphicx}

\begin{document}
\title{Magnetization structure of a Bloch point singularity}
\author{Ricardo Gabriel Elías}\email{gabriel.elias@im2np.fr}
\author{Alberto Verga}\email{Alberto.Verga@univ-provence.fr}
   \affiliation{%
   Aix-Marseille Université, IM2NP, 
   Campus de St Jérôme, Case 142, 13397 Marseille, France}
   \altaffiliation{%
    IM2NP, CNRS-UMR 6242.
   }

\date{\today}

\begin{abstract}
Switching of magnetic vortex cores involves a topological transition characterized by the presence of a magnetization singularity, a point where the magnetization vanishes (Bloch point). We analytically derive the shape of the Bloch point that is an extremum of the free energy with exchange, dipole and the Landau terms for the determination of the local value of the magnetization  modulus.
\end{abstract}


\maketitle

The interest in the dynamics of magnetic vortices was renewed by the discovery of fast core reversal by a varying external excitation (magnetic field\cite{Van-Waeyenberge-2006fk} or spin current\cite{Yamada-2007qt}). Micromagnetic simulations of vortex core switching\cite{Hertel-2007jy} revealed that the underlying mechanism, the annihilation of a vortex-antivortex pair\cite{Hertel-2006ib}, needs the mediation of a magnetization singularity: a magnetic monopole or Bloch point.\cite{Thiaville-2003rq} This magnetization structure was first studied by Feldtkeller,\cite{Feldtkeller-1965uq} 
who showed that it is mainly determined by the exchange energy; later on Döring\cite{Doring-1968bf} calculated this specific energy and showed that its value would be a topologically invariant. He considered a family of magnetization textures differing in their local rotation angle $\gamma$ (with respect to the radial direction) and found that minimization of the demagnetization energy density selected a specific angle $\gamma\approx 112^\circ$. However, any approach within the micromagnetic approximation (the magnetization strength is at its saturation value) cannot account for the internal structure of the singularity that imposes the vanishing of the magnetization vector. In order to investigate the region near the singular point Galkina et al.\cite{Galkina-1993uf} included the Landau magnetic energy, although they neglected the demagnetization term, and showed that the magnetization vector modulus increases linearly with the radial distance from the origin. Therefore, to understand the topological transitions between different vortex states, for which magnetic monopoles are required,\cite{Tretiakov-2007fc} it is important to go beyond the micromagnetic approximation. In this paper we compute the magnetization field of a Bloch point taking into account the exchange, Landau and demagnetizing energies. We obtain two solutions, the first one, corresponding to a local minimum of the energy density, is characterized by a linear magnetization modulus near the origin and by an essentially \emph{azimuthal} magnetization configuration with a rotation angle $\gamma$ (incidentally rather close to the one found by Döring\cite{Doring-1968bf}); the second one, also linear near the center, but with a \emph{radial} magnetization vector (hedgehog Bloch point), is valid over a finite spherical region, and corresponds to a local maximum of the energy density.


In order to determine the magnetization field $\bm{M}(\bm{r})$ of a Bloch point in a ferromagnetic nanostructure we consider the free energy $\mathcal{F}=\mathcal{F}[\bm{M},\Phi]$ as a functional of $\bm{M}$ and of the magnetic potential $\Phi$,
\begin{equation}
\label{Fdim}
\mathcal{F}=\mathcal{F}[\bm{M},\Phi]=\int dV\bigg[\frac{A}{2}(\nabla\bm{M})^2+f_\mathrm{L}(M)\bigg]+\mathcal{F}_H,
\end{equation}
where $dV$ is the volume element, $A$ is the exchange energy constant, $f_\mathrm{L}$ is the Landau energy density
\begin{equation}
\label{fLdim}
f_\mathrm{L}(M)=aM^2+bM^4,
\end{equation}
$a=a(T)$ is in general a function of the temperature, $a<0$ in the relevant ferromagnetic state, $b$ is a dimensional constant, and the energy of the demagnetizing field
\begin{equation}
\label{FHdim}
\mathcal{F}_H=\mathcal{F}_H[\bm{M},\Phi]=-\mu_0\int dV\bigg(\bm{M}\cdot\bm{H}+\frac{H^2}{2}\bigg),
\end{equation}
with $\bm{H}(\bm{r})=-\nabla\Phi(\bm{r})$ the magnetic field. We introduce the following units: length, $\ell=(A/\mu_0)^{1/2}$; magnetization, $\mathcal{M}_s=(-a/2b)^{1/2}$; and energy, $\mathcal{E}=(-aA/2b)\ell$. In this units system the free energy becomes,  
\begin{equation}
\label{F}
\mathcal{F}=\int dV\bigg[\frac{1}{2}(\nabla\bm{M})^2+\nu f(M)+\bm{M}\cdot\nabla\Phi-\frac{|\nabla\Phi|^2}{2}\bigg],
\end{equation}
where $f(M)=-M^2+M^4/2$, and $\nu=|a|/\mu_0$ is the only nondimensional parameter of the system; it can be written as
\begin{equation}
\nu=\ell^2/\ell_0^2,\;\ell=(A/\mu_0)^{1/2},\;
\ell_0=(-A/a)^{1/2},
\end{equation}
where $\ell$ is related to the exchange length $\ell_A$ ($\ell_A=\sqrt{2}\ell$) and $\ell_0$ is the characteristic length of the magnetization intensity $M=|\bm{M}|$ variation, as will be demonstrated below. 


The equilibrium distributions of the magnetic potential and the magnetization field are determined by the variation of $\mathcal{F}$ with respect to $\Phi$, and $\bm{M}$. The variational derivative of (\ref{F}) with respect to $\Phi$, leads to the Maxwell equations,
\begin{equation}
\label{lapphi}
\nabla^2\Phi=\nabla\cdot\bm{M},
\end{equation}
in the magnetic domain, and, at the surface boundary
\begin{equation}
\label{nm}
\hat{\bm{n}}\cdot\bm{M}=\hat{\bm{n}}\cdot\Delta\bm{H},
\end{equation}
where $\hat{\bm{n}}$ is the normal and $\Delta\bm{H}$ the discontinuity of the magnetic field. The variational derivative of (\ref{F}) with respect to $\bm{M}$ leads to 
\begin{equation}
\label{eqM}
-\nabla^2\bm{M}+\nu\frac{\partial f(M)}{\partial \bm{M}}+\nabla\Phi=0.
\end{equation}
This equation can be transformed into an integro-differential equation for the magnetization field using the explicit solution of (\ref{lapphi}-\ref{nm}) in terms of $\bm M$,
\begin{equation}
\label{Phi}
\Phi(\bm{r})=-\frac{1}{4\pi}
\int_{\mathcal{V}} dV'\frac{\nabla'\cdot\bm{M}'}{|\bm{r}-\bm{r}'|}+
\frac{1}{4\pi}
\int_{\partial\mathcal{V}} \frac{d\bm{S}'\cdot\bm{M}'}{|\bm{r}-\bm{r}'|}
\end{equation}
where prime variables refer to the magnetic domain $\mathcal{V}$ and its boundary $\partial\mathcal{V}$.


The problem now is to determine the structure of the Bloch point as a particular solution of Eq.~(\ref{eqM}). This can be done by introducing an appropriate ansatz. To compute the demagnetizing field we consider a simple geometry, we take for $\mathcal{V}$ a sphere of radius $R$; in this region, we choose a magnetization field that generalizes the Feldtkeller\cite{Feldtkeller-1965uq} and Döring\cite{Doring-1968bf} ansatz, adding a $r$-dependent magnetization modulus,
\begin{equation}
\label{ansatz}
\bm{M}=M(r)\bm{m}_\gamma(\theta,\phi)
\end{equation}
where the unit vector $\bm{m}_\gamma$ has the topology of a Bloch point and satisfies the condition to be an extremum of the exchange free energy (neglecting other terms in $\mathcal{F}$). The simplest one-parameter solution, depending on a rotation angle $\gamma$, can be written as,\cite{Doring-1968bf}
$$
\bm{m}_\gamma =(\cos(\phi+\gamma)\sin\theta,\sin(\phi+\gamma)\sin\theta,\cos\theta),
$$ 
in cartesian coordinates $(\hat{\bm{x}},\hat{\bm{y}},\hat{\bm{z}})$, where $\theta$, $\phi$ are, simultaneously, spherical angles for both $\hat{\bm{r}}$ and $\bm{m}_\gamma$. In the following calculation we use the notations
\[
c_\pm(\gamma)=\frac{1}{2}(1\pm \cos\gamma),\;
c_2(\gamma)=\frac{1}{3}(1+2 \cos\gamma)
\]
and represent vectors in spherical coordinates $(\hat{\bm{r}},\hat{\bm{\theta}},\hat{\bm{\phi}})$:
\begin{equation}
\bm{m}_\gamma =\left(\begin{array}{c}
c_+(\gamma)+c_-(\gamma)\cos 2\theta\\
-c_-(\gamma) \sin2 \theta\\
\sin \gamma  \sin \theta
\end{array}\right),
\end{equation}
where $\gamma=0$ corresponds to the hedgehog configuration, and $\gamma=\arccos(-11/29)$ to the spiral Bloch point found by Döring.\cite{Doring-1968bf}


The magnetization intensity vanishes near the singularity and reaches its saturation value far from the origin (at a distance $r\gg \ell_0/\ell$). Therefore, a solution of (\ref{eqM}) must satisfy:
\begin{equation}
\label{lim}
\lim_{r\rightarrow 0}M(r)\rightarrow 0,\;
\lim_{r\rightarrow \infty}M(r)\rightarrow M_s(\nu),
\end{equation}
where the saturation magnetization $M_s$ is in general a function of $\nu$. Inserting the ansatz (\ref{ansatz}) into (\ref{eqM}) one obtains the integro-differential equation,
\begin{equation}
\label{MH}
\left[\hat{D}M+2\nu \left(M-M^3\right)\right]\bm m_\gamma +\bm{H}=0,
\end{equation}
where
\begin{equation}
\hat{D}=\frac{\partial^2}{\partial r^2}+\frac{2}{ r} \frac{\partial}{\partial r}-\frac{2}{ r^2},
\end{equation}
is the radial part of the laplacian with angular momentum $l=1$.


To find the demagnetizing potential we insert the ansatz (\ref{ansatz}) into the equation for the potential (\ref{Phi}), and note that the magnetization volume $\nabla\cdot\bm{M}$ and surface $\hat{\bm{r}}\cdot\bm{M}$ charges, can be expressed as an expansion over the spherical harmonics $Y_0^0(\theta,\phi)$ and $Y_2^0(\theta,\phi)$:
\begin{equation}
-\Phi(\bm{r})=2\sqrt{\pi}c_2(\gamma) F(r) Y_0^0 +
\frac{8\sqrt{\pi}}{3\sqrt{5}}c_-(\gamma) G(r) Y_2^0
\end{equation}
where we define the functions of the radial coordinate $r$,
\begin{align}
F(r)=&\frac{1}{r}\int_0^r 
\left(\frac{\partial M'}{\partial r'}+\frac{2 M'}{r'}\right)r'^2 dr'+\nonumber\\
&\int_r^R \left(\frac{\partial M'}{\partial r'}+
 \frac{2 M'}{r'}\right) r'dr'-\frac{M(R)R^2}{r_>} ,
\end{align}
and
\begin{align}
G(r)=&\frac{1}{5r^3}
\int_0^r \left(\frac{\partial M'}{\partial r'}-\frac{M'}{r'}\right)r'^4dr'+\nonumber\\
&\frac{r^2}{5}\int_r^R \left(\frac{\partial M'}{\partial r'}-\frac{M'}{r'}\right) \frac{dr'}{r'}-\frac{M(R)R^2 r_<^2}{5r_>^3}.
\end{align}
(We use the standard notation $r_>=r'$ and $r_<=r$ if $r<r'$, etc.) From this formula one immediately deduces the demagnetizing field,
\begin{equation}
\bm H=\left(\begin{array}{c}
c_2(\gamma)F'(r)+c_-(\gamma)(\frac{1}{3}+\cos2\theta)G'(r)\\
-\frac{2}{r}c_-(\gamma)G(r)\sin2\theta\\
0
\end{array}\right)
\end{equation}
(in a spherical frame), that can be used in Eq.~(\ref{MH}) to obtain the following four equations:
\begin{align}
\sin\gamma \,O(M)&=0, \label{one}\\
c_-(\gamma)\left[O(M)+\frac{2}{r}G(r)\right]&=0, 
   \label{two}\\
c_+(\gamma)O(M)-c_2(\gamma)F'(r)+\frac{c_-(\gamma)}{3}G'(r)&=0, 
   \label{three}\\
c_-(\gamma)\left[O(M)+G'(r) \right]&=0 \label{four},
\end{align}
where
\begin{equation}
O(M)=\hat{D}M(r)+2\nu \big[M(r)-M(r)^3\big].
\end{equation}
Different magnetization textures can be solution of these equations depending on their characteristic length scales and possessing different energies. The existence of the parameter $\nu$ (usually large for ferromagnetic materials) allows the separation of three regions: a singular core region $r<\ell_0/\ell$, characterized by a rapid variation of the magnetization modulus; an intermediate region (the micromagnetic core) $\ell_0/\ell<r<1$; and an external region $r>1$, dominated by the dipolar energy. In the singular core region ($r\lesssim1$), where one can formally take $\nu\rightarrow\infty$, the demagnetizing terms are negligible compared to the exchange and Landau ones. It is then possible to distinguish between two cases: (i) $\sin\gamma \ne 0$, implying $O(M)=0$, that leads to a \emph{local} solution, valid near the center of the Bloch point, and, as we shall see, corresponding to a minimum of the energy; and (ii) $\sin\gamma=0$, that allows for a \emph{global} solution, corresponding to a maximum of the energy.


\begin{figure}
\centering
\includegraphics[width=.48\textwidth]{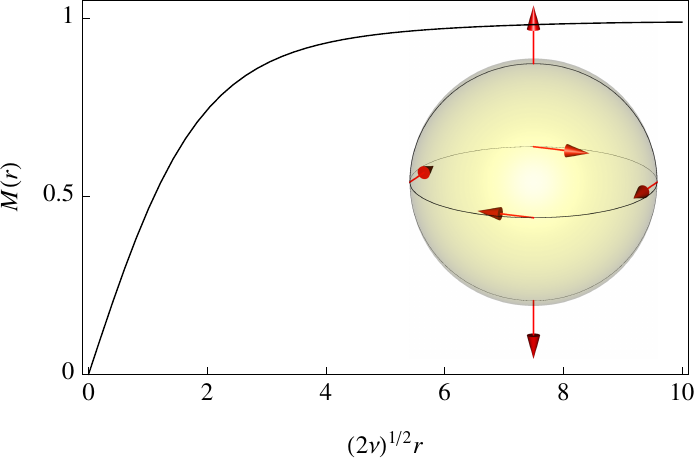}\\
\includegraphics[width=.48\textwidth]{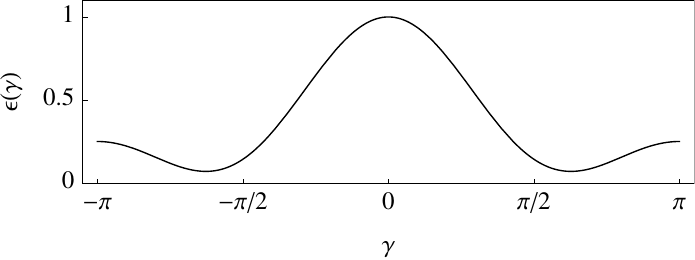}
\caption{(Color online) Bloch point for the minimun energy structure, $\gamma\ne0$. (top) Magnetization intensity, in units of $\mathcal{M}_s$ as a function of the stretched coordinate $\sqrt{2\nu}r$ in units of $\ell$, and its angular distribution (inset) for $\gamma=113^\circ$. (bottom) Nondimensional demagnetization energy  as a function of the rotation angle $\gamma$.  }
\label{i}
\end{figure}


First we consider case (i), the minimum energy local solution. From (\ref{one}), the assumption $\sin\gamma \ne 0$, implies $O(M)=0$, and compatibility with the other equations is possible in the singular region, where the demagnetizing field is small. We have thus, to find a solution of
\begin{equation}
\label{eqi}
O(M)=\hat{D}M(r)+2\nu\big[M(r)-M(r)^3\big]=0,
\end{equation}
with the boundary conditions (\ref{lim}). A simple scaling transformation $r\rightarrow (2\nu)^{1/2}r$ allows to scale out the parameter $\nu$. This implies that the function $M=M((2\nu)^{1/2}r)$ is universal, which can be considered as a generalization of the invariance statement by Döring\cite{Doring-1968bf}. Let us consider first the singular region $r<\ell_0/\ell\ll1$ where one can assume that $M(r)=C(\nu)r$ is a linear function of the radial coordinate. This choice is motivated by the property that a linear magnetization satisfies $\hat{D}Cr=0$, and is then an asymptotic solution of (\ref{eqi}). The constant $C$ depends trivially on the parameter $\nu$; using the above radial scaling, one finds $C(\nu)= (2\nu)^{1/2}C_\infty$ where the universal constant $C_\infty$ must be determined as an eigenvalue of $O(M)=0$ satisfying (\ref{lim}). It is important to note that a linear magnetization amplitude near $r=0$ implies $G\sim F\sim r^2$, meaning \emph{a posteriori} that the demagnetizing terms in (\ref{two}-\ref{four}) are indeed negligible compared to the exchange and Landau terms in the singular region. We consider now the core region $r\lesssim1$, where, in accordance with (\ref{lim}), the modulus $M$ approaches the saturation value $M=1$, canceling the Landau term in (\ref{eqi}). We remark that for the core region $r\approx 1$, the large $\nu$ condition ensures that $M$ reaches its saturation value and that the demagnetization terms are also negligible. Indeed, the characteristic length for which the magnetization saturates may be estimated by $(2\nu)^{1/2}r\approx \mathcal{O}(1)$, or $r\approx \mathcal{O}(\ell_0/\ell)$ that usually is very small. Therefore, the solution of (\ref{eqi}) is consistent with the system (\ref{one}-\ref{four}) throughout the core region. 


The universal magnetization profile, a solution of $O(M)=0$ satisfying (\ref{lim}) with $M_s\rightarrow 1$, was computed numerically and is shown in Fig.~\ref{i}. We observe that the magnetization saturates for values of $(2\nu)^{1/2}r\approx 4$; as a consequence the validity condition of the solution is well verified for $\nu>10$. The numerical profile is also consistent with the linear magnetization near the center with $C_\infty=0.506$, and a saturation $M=1$ for large $r$. It is worth noticing that in this case the value of $\gamma$ is not determined, showing that the ansatz (\ref{ansatz}) represents a one-parameter family of solutions for the internal structure of the Bloch point. However, the demagnetizing field, although negligible for the determination of the magnetization radial profile, should select a specific value of $\gamma$ in order to minimize the free energy.\cite{Doring-1968bf} The relevant part of the free energy is the demagnetizing field density (the Landau and exchange terms are independent of $\gamma$),
\begin{equation}
-( \bm M \cdot \bm H+H^2/2)dV=(H^2/2)dV.
\end{equation}
In the inner region, where the magnetization is given by 
\begin{equation}
M(r)=(2\nu)^{1/2}C_\infty r,
\end{equation}
the demagnetizing field writes
\begin{equation}
\bm H=Cr\left(\begin{array}{c}
-c_2(\gamma)+c_-(\gamma)(\frac{1}{3}-\frac{2}{5}\cos2\theta)\\
\frac{2}{5}c_-(\gamma)\sin2\theta\\
0
\end{array}\right).
\end{equation}
After integration over the spherical angles, the part of the energy density depending explicitly on $\gamma$, is
\begin{equation}
\epsilon(\gamma)=(28 \cos \gamma +18 \cos 2 \gamma +29)/75,
\end{equation}
(see Fig.~\ref{i} bottom panel) whose minimization gives $\gamma=\arccos(-7/18)\approx 113^\circ$. The distribution of the magnetization in this case is represented in the Fig.~\ref{i} (inset). Extending this argument to the region $M\sim 1$, one finds the Döring result $\gamma=\arccos(-11/29)\approx 112^\circ$, showing that the rotation angle $\gamma$ must actually be a function of the radial coordinate, although its slow variation confirms the approximated validity of the ansatz (\ref{ansatz}) for this structure.


\begin{figure}
\centering
\includegraphics[width=.48\textwidth]{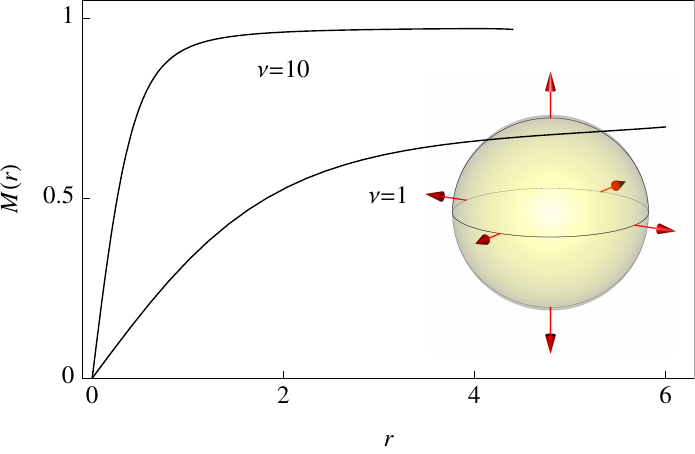}\\
\includegraphics[width=.48\textwidth]{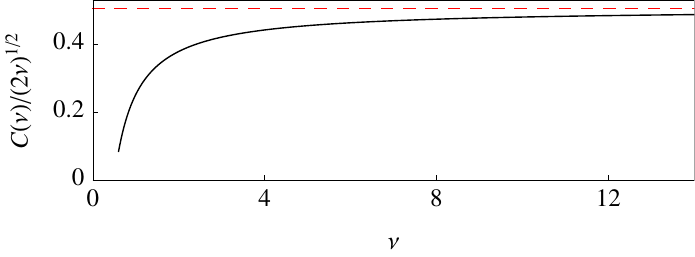}
\caption{(Color online) Bloch point for the maximum energy structure, $\gamma=0$. (top) Nondimensional magnetization intensity profile for $\nu=1, 10$, and its angular distribution (inset). (bottom) Variation of the normalized slope at the origin as a function of $\nu$, and its asymptote $C_\infty$ (dashed line).}
\label{ii}
\end{figure}

Second, we consider case (ii), the global solution with $\sin\gamma=0$, corresponding to a maximum of the energy structure. In this case we note that Eqs.~(\ref{one}), (\ref{two}), and (\ref{four}) hold identically, and then we are left to solve
\begin{equation}
\label{eqii}
\hat{D}M(r)+(2\nu-1)M(r)-2\nu M(r)^3=0
\end{equation}
where the last term comes from $F'(r)=-M(r)$ in the magnetized domain $r<R$. The behavior of $M(r)$ for $\gamma=0,\pi$, shown in Fig.~\ref{ii}, is qualitatively similar to the $\gamma\ne 0$ case but with a saturation magnetization,
\begin{equation}
M_s(\nu)=\sqrt{(2\nu-1)/2\nu},
\end{equation}
smaller than the $\nu\rightarrow\infty$ limit value of $1$, relevant in the minimum energy case, where we neglected the demagnetizing field. It is worth noting, that in spite of the similarity between (\ref{eqi}) and (\ref{eqii}), the two cases are completely different: the global solution provide an exact solution of the magnetization profile in the spherical region. It is not possible to connect the two solutions, the local solution corresponds to the minimum of $\epsilon(\gamma)$ while global case to its maximum. Incidentally, we remark that the second (\ref{lim}) condition is violated for $\nu<1/2$, showing that for small $\nu$ the radial Bloch point does not exist as a stationary state. Near the origin the magnetization is linear with a slope $C(\nu)$, represented in Fig.~\ref{ii}, that tends to $C_\infty(2\nu)^{1/2}$ for large $\nu$. This slope determines the typical size of the singular core,
\begin{equation}
\label{l}
\ell_s=\ell/C(\nu)
\end{equation}
which tends to zero when $\nu\rightarrow\infty$, showing the singular behavior near the origin in the micromagnetic approximation. The bottom panel of Fig.~\ref{ii} shows the slope $C$, whose inverse determines the characteristic length of the singular core (\ref{l}), normalized to $(2\nu)^{1/2}$.

In conclusion, we have shown that the interplay of exchange, dipolar and Landau energies, determines the internal structure of the Bloch point, beyond the micromagnetic approximation. We have generalized the pure radial solution of Galkina,\cite{Galkina-1993uf} and the constant magnetization modulus solutions derived from the original Feldtkeller\cite{Feldtkeller-1965uq} ansatz. Noteworthy, the actual structure of a Bloch point eventually depends on the boundary conditions, that will determine the demagnetizing field and, near the singularity, select the rotation angle $\gamma$ (in general a function of the radius).\cite{Jourdan-2008lq} An estimation of the size of the inner core gives $\ell_s\approx 2^{1/2}\ell_0\sim 1\,\mathrm{nm}$ (for large $\nu$ and typical permalloy parameters values, $\ell\approx 10 \ell_0$), shows that in order to resolve the dynamics it is necessary to introduce quantum effects,\cite{Miltat-2002kx} at least in a semiclassical approximation. In fact, such effects will control the dissipation processes that lead to the actual topological transition.


%

\end{document}